\documentclass[12pt]{article}
\title{\centerline
\bf Solution of the Dirac equation in presence of an 
uniform magnetic field}
\bigskip
\author{Kaushik Bhattacharya
\thanks{e-mail
addresses:kaushik@nucleares.unam.mx,
}\\
\normalsize
Instituto de Ciencias Nucleares,\\
\normalsize
Universidad Nacional Autonoma de Mexico,\\
\normalsize
Circuito Exterior, C.U., A. Postal 70-543, C.P. 04510 Mexico D.F.,\\
\normalsize
Mexico.
}
\begin{document}
\newcommand{\Tr}{\mathop{\rm Tr}\nolimits}
\newcommand{\para}{_\parallel}
\newcommand{\pr}{_\perp}
\newcommand{\fs}{\rlap/}
\def\twidle{\widetilde}
\def\f{\frac}
\def\omit#1{_{\!\rlap{$\scriptscriptstyle \backslash$}
{\scriptscriptstyle #1}}}
\def\vec#1{\mathchoice 
	{\mbox{\boldmath $#1$}}
	{\mbox{\boldmath $#1$}}
	{\mbox{\boldmath $\scriptstyle #1$}}
	{\mbox{\boldmath $\scriptscriptstyle #1$}}
}
\def\eqn#1{Eq.\ (\ref{#1})}
\maketitle
\begin{abstract}
In this work we discuss the properties of the solutions of the Dirac
equation in presence of an uniform background magnetic field. In
particular we focus on the nature of the solutions, their
ortho-normality properties and how these solutions depend on the
choice of the vector potential giving rise to the magnetic field. We
explicitly calculate the spin-sum of the solutions and using it we
calculate the propagator of the electron in presence of an uniform
background magnetic field.  
\end{abstract}
\section{Introduction}
\label{int}
Calculations of elementary particle decays and scattering
cross-sections in presence of a background magnetic field are commonly
found in literature \cite{Can69, MOc69, MOc70,
Dorofeev:az,Dicus:2007gb}. These calculations became more important
after it was understood that the neutron star cores can sustain
magnetic fields of the order of $10^{13}\,{\rm G}$ or more. These
realistic fields may be very complicated in their structure but for
simplicity many of the times we assume these fields to be uniform. The
advantage of an uniform magnetic field is that in presence of this
field the Dirac equation can be exactly solved. Once the the Dirac
equation is exactly solved then we can proceed to quantize those
solutions and calculate elementary particle decays and scattering
cross-sections in presence of the background magnetic field. In this
article we will solve the Dirac equation in a background magnetic
field and discuss about the nature of the solutions. We will quantize
the fermionic theory in presence of a magnetic field and will end with
the calculation of the electron propagator. As the calculations rely
heavily on the choice of the background gauge field giving rise to the
magnetic field we will discuss about the gauge dependence of the
various quantities calculated in this article and infer about the
gauge invariance of physical quantities as scattering cross-sections
and decay rates calculated in presence of a magnetic field.
\section{Charged fermion in presence of a magnetic field}
\label{dsol}
In this article we will assume that the uniform classical background
magnetic field is along the $z$-direction of the co-ordinate axis. The
background gauge fields giving rise to a magnetic field along the
$z$-direction, of magnitude ${\mathcal B}$, can be fixed in many ways:
\begin{eqnarray}
A^0_{\rm B} = A^y_{\rm B} = A^z_{\rm B} = 0 \,, \qquad 
A^x_{\rm B} = -y{\mathcal B} + b\,.
\label{GA}
\end{eqnarray}
or 
\begin{eqnarray}
A^0_{\rm B} = A^x_{\rm B} = A^z_{\rm B} = 0 \,, \qquad 
A^y_{\rm B} = x{\mathcal B}+c\,.
\label{GB}
\end{eqnarray}
or
\begin{eqnarray}
A^0_{\rm B} = A^z_{\rm B} = 0 \,, \qquad 
A^y_{\rm B} = \frac{1}{2} x{\mathcal B}+d\,, \qquad
A^x_{\rm B} = -\frac{1}{2} y{\mathcal B}+g\,,
\label{GC}
\end{eqnarray}
where $b$, $c$, $d$ and $g$ are constants. Here $A^\mu_{\rm B}$
designates that the gauge field is a classical background field and
not a quantized dynamical field. In the above equations $x$, $y$ are
just coordinates and not 4-vectors. In this article we will assume
that the gauge configuration as given in Eq.~(\ref{GA}) with
$b=0$. More over in this article we will be employing the Dirac-Pauli
representation of the Dirac matrices.
\subsection{The solution of the Dirac equation}
The Dirac equation for a particle of mass $m$ and charge $eQ$, in
presence of a magnetic field is given by:
\begin{eqnarray}
i {\partial\psi \over \partial t} = {\mathcal H}_{\rm B}\,\psi \,,
\label{DiracEq}
\end{eqnarray}
where the Dirac Hamiltonian in presence of a magnetic field is given by:
\begin{eqnarray}
{\mathcal H}_{\rm B}= \vec \alpha \cdot {\bf \Pi} + \beta m\,.
\end{eqnarray}
Here $\Pi^\mu$ is the kinematic momentum of the charged fermion. In
our convention, $e$ is the positive unit of charge, taken as usual to
be equal to the proton charge.  From Eq.~(\ref{DiracEq}) we can infer
that for the stationary states, we can write:
\begin{eqnarray}
\psi = e^{-iEt} \left( \begin{array}{c} \phi \\ \chi \end{array}
\right) \,,
\end{eqnarray}
where $\phi$ and $\chi$ are 2-component objects. With this notation, we can 
write Eq.~(\ref{DiracEq}) as:
\begin{eqnarray}
(E-m)\phi &=& \vec \sigma \cdot (-i\vec\nabla - eQ\vec A) \chi \,, 
\label{eq1}\\*
(E+m) \chi &=& \vec \sigma \cdot (-i\vec\nabla - eQ\vec A) \phi \,.
\label{eq2}
\end{eqnarray}
Eliminating $\chi$, we obtain 
\begin{eqnarray}
(E^2 - m^2)\phi &=& \Big[ \vec \sigma \cdot (-i\vec\nabla - eQ\vec A) 
\Big]^2 \phi \,.
\label{phieq1} 
\end{eqnarray}
With our choice of the vector potential, Eq.\ (\ref{phieq1}) reduces
to the form
\begin{eqnarray}
(E^2 - m^2)\phi 
&=& \Big[ -\vec\nabla^2 + (eQ{\mathcal B})^2 y^2 - eQ{\mathcal B}(2iy
{\partial\over 
\partial x} + \sigma_3) \Big] \phi \,. 
\label{phieq} 
\end{eqnarray}
Here $\sigma_3$ is the diagonal Pauli matrix. Noticing that the
co-ordinates $x$ and $z$ do not appear in the equation except through
the derivatives, we can write the solutions as
\begin{eqnarray}
\phi = e^{i \vec {\scriptstyle p} \cdot \vec{\scriptstyle X} \omit y}
f(y) \,, 
\label{phiform}
\end{eqnarray}
where $f(y)$ is a 2-component matrix which depends only on the
$y$-coordinate, and possibly some momentum components, as we will see
shortly. We have also introduced the notation $\vec X$ for the spatial
co-ordinates (in order to distinguish it from $x$, which is one of the
components of $\vec X$), and $\vec X\omit y$ for the vector $\vec X$
with its $y$-component set equal to zero. In other words, $\vec p\cdot
\vec X{\omit y} \equiv p_xx+p_zz$, where $p_x$ and $p_z$ denote the
eigenvalues of momentum in the $x$ and $z$ directions.\footnote{It is
to be understood that whenever we write the spatial component of any
vector with a lettered subscript, it would imply the corresponding
contravariant component of the relevant 4-vector.}

There will be two independent solutions for $f(y)$, which can be
taken, without any loss of generality, to be the eigenstates of
$\sigma_3$ with eigenvalues $s=\pm 1$. This means that we choose the
two independent solutions in the form
\begin{eqnarray}
f_+ (y) = \left( \begin{array}{c} F_+(y) \\ 0 \end{array} \right) \,,
\qquad 
f_- (y) = \left( \begin{array}{c} 0 \\ F_-(y) \end{array} \right) \,.
\end{eqnarray}
Since $\sigma_3 f_s = sf_s$, the differential equations satisfied by
$F_s$ is
\begin{eqnarray}
{d^2F_s \over dy^2} - (eQ{\mathcal B}y + p_x)^2 F_s + (E^2 - m^2 -
p_z^2 + eQ{\mathcal B}s) 
F_s = 0 \,,
\label{Fseqn}
\end{eqnarray}
which is obtained from Eq.\ (\ref{phieq}).  The solution is obtained
by using the dimensionless variable
\begin{eqnarray}
\xi = \sqrt{e |Q| {\mathcal B}} \left( y + {p_x \over eQ{\mathcal B}}
\right) \,, 
\label{xi}
\end{eqnarray}
which transforms Eq.\ (\ref{Fseqn}) to the form
\begin{eqnarray}
\left[ {d^2 \over d\xi^2} -\xi^2 + a_s \right] F_s = 0 \,,
\label{diffeqn}
\end{eqnarray}
where
\begin{eqnarray}
a_s = {E^2 - m^2 - p_z^2 + eQ{\mathcal B}s \over e|Q|{\mathcal B}} \,.
\label{impb0}
\end{eqnarray}
This is a special form of Hermite's equation, and the solutions exist
provided $a_s=2\nu+1$ for $\nu=0,1,2,\cdots$. This provides the energy
eigenvalues 
\begin{eqnarray}
E^2 = m^2 + p_z^2 + (2\nu+1)e|Q|{\mathcal B} - eQ{\mathcal B}s \,,
\label{E}
\end{eqnarray}
and the solutions for $F_s$ are
\begin{eqnarray}
N_{\nu} e^{-\xi^2/2} H_{\nu}(\xi) \equiv I_{\nu}(\xi) \,,
\label{In}
\end{eqnarray}
where $H_\nu$ are Hermite polynomials of order $\nu$, and $N_\nu$ are
normalizations which we take to be
\begin{eqnarray}
N_\nu = 
\left( {\sqrt{e|Q|{\mathcal B}} \over \nu! \, 2^\nu \sqrt{\pi}} \,
\right)^{1/2} \,.  
\label{Nn}
\end{eqnarray}
With our choice, the functions $I_\nu$ satisfy the completeness
relation
\begin{eqnarray}
\sum_\nu I_\nu(\xi) I_\nu(\xi_\star) = \sqrt{e|Q|{\mathcal B}} \;
\delta(\xi-\xi_\star) = \delta (y-y_\star) \,, 
\label{completeness}
\end{eqnarray}
where $\xi_\star$ is obtained by replacing $y$ by $y_\star$ in Eq.\
(\ref{xi}).

So far, $Q$ was arbitrary.  We now specialize to the case of
electrons, for which $Q=-1$.  The solutions are then conveniently
classified by the energy eigenvalues
\begin{eqnarray}
E_n^2 = m^2 + p_z^2 + 2ne{\mathcal B} \,,
\label{En}
\end{eqnarray}
which is the relativistic form of Landau energy levels. The solutions
are two fold degenerate in general: for $s=1$, $\nu=n-1$ and for
$s=-1$, $\nu=n$. In the case of $n=0$, from Eq.\ (\ref{E}) we see that
for $Q=-1$, $\nu=-\frac{1}{2}(1+s)$, and as $\nu$ cannot be negative
$s=-1$.  Thus the $n=0$ state is not degenerate. The solutions can have
positive or negative energies. We will denote the positive square root
of the right side by $E_n$. Representing the solution corresponding to
this $n$-th Landau level by a superscript $n$, we can then write for
the positive energy solutions,
\begin{eqnarray}
f_+^{(n)} (y) = \left( \begin{array}{c} 
I_{n-1}(\xi) \\ 0 \end{array} \right) \,,
\qquad 
f_-^{(n)} (y) = \left( \begin{array}{c} 
0 \\ I_n (\xi) \end{array} \right) \,.
\label{fsolns}
\end{eqnarray}
For $n=0$, the solution $f_+$ does not exist. We will consistently
incorporate this fact by defining
\begin{eqnarray}
I_{-1} (y) = 0 \,,
\label{I_-1}
\end{eqnarray}
in addition to the definition of $I_n$ in Eq.\ (\ref{In}) for
non-negative integers $n$.

The solutions in Eq.\ (\ref{fsolns}) determine the upper components of
the spinors through Eq.\ (\ref{phiform}). The lower
components, denoted by $\chi$ earlier, can be solved using
Eq.\ (\ref{eq2}), and finally the positive energy solutions of the
Dirac equation can be written as
\begin{eqnarray}
e^{-ip\cdot X {\omit y}} U_s (y,n,\vec p \omit y) \,,
\label{psol}
\end{eqnarray}
where $X^\mu$ denotes the space-time coordinate. And $U_s$ are given
by \cite{Bhattacharya:2002qf, Bhattacharya:2004nj, Bhattacharya:2002aj}
\begin{eqnarray}
U_+ (y,n,\vec p \omit y) = \left( \begin{array}{c} 
I_{n-1}(\xi) \\[2ex] 0 \\[2ex] 
{\strut\textstyle p_z \over \strut\textstyle E_n+m} I_{n-1}(\xi) \\[2ex]
-\, {\strut\textstyle \sqrt{2ne{\mathcal B}} \over \strut\textstyle 
E_n+m} I_n (\xi) 
\end{array} \right) \,, \qquad 
U_- (y,n,\vec p \omit y) = \left( \begin{array}{c} 
0 \\[2ex] I_n (\xi) \\[2ex]
-\, {\strut\textstyle \sqrt{2ne{\mathcal B}} \over \strut\textstyle E_n+m}
I_{n-1}(\xi) \\[2ex] 
-\,{\strut\textstyle p_z \over \strut\textstyle E_n+m} I_n(\xi) 
\end{array} \right) \,. 
\label{Usoln}
\end{eqnarray}

For the case of positrons which are positively charged negative energy
solutions of the Dirac equation we have to put $Q=-1$ and in this case
also we can write their energy as in Eq.~(\ref{En}) but this time
Eq.~(\ref{E}) predicts that the $n=0$ solution must only have the
$s=1$ component. Although the dispersion relation of the electrons and
positrons become different in presence of a magnetic field but they
can be written in a unique form as given in Eq.~(\ref{En}), the
difference shows up in the spin of the zeroth Landau level state. A
similar procedure, as used for solving for the positive energy
spinors, can be adopted to solve for the negative energy spinors and
the solutions are:
\begin{eqnarray}
e^{ip\cdot X{\omit y}} V_s (y,n, \vec p\omit y) \,,
\label{nsol}
\end{eqnarray}
where 
\begin{eqnarray}
V_- (y,n,\vec p\omit y) = \left( \begin{array}{c} 
{\strut\textstyle p_z \over \strut\textstyle E_n+m}
I_{n-1}(\widetilde\xi) \\[2ex] 
{\strut\textstyle \sqrt{2ne{\mathcal B}} \over \strut\textstyle E_n+m} 
I_n (\widetilde\xi)  \\[2ex] 
I_{n-1}(\widetilde\xi) \\[2ex] 0
\end{array} \right) \,, \qquad 
V_+ (y,n,\vec p\omit y) = \left( \begin{array}{c} 
{\strut\textstyle \sqrt{2ne{\mathcal B}} \over \strut\textstyle E_n+m}
I_{n-1}(\widetilde\xi) \\[2ex] 
-\,{\strut\textstyle p_z \over \strut\textstyle E_n+m}
I_n(\widetilde\xi)  \\[2ex] 
0 \\[2ex] I_n (\widetilde\xi)
\end{array} \right) \,.
\label{Vsoln}
\end{eqnarray}
where $\widetilde\xi$ is obtained from $\xi$ by changing the sign of
the $p_x$-term. These solutions are eigenstates of $\Pi_x$ and $\Pi_z$
but not of $\Pi_y$. As $\Pi_x$ and $\Pi_y$ do not commute we cannot
have simultaneous eigenstates of both.

The solutions of the Dirac equation in presence of a magnetic field
are exact solutions and not perturbative excitations around the free
Dirac equation solutions, which is evident from Eq.\ (\ref{impb0}).
Consequently we cannot put ${\mathcal B} \to 0$ in the final
solutions, in Eq.~(\ref{Usoln}) and Eq.~(\ref{Vsoln}), and expect we
will get back the free Dirac solutions. Mathematically in the zero
field limit the quantization condition in Eq.\ (\ref{impb0}) fails and
in that limit the solutions of Eq.~(\ref{diffeqn}) becomes
indeterminate. Physically we can say that the solutions in
Eq.~(\ref{Usoln}) and Eq.\ (\ref{Vsoln}) are specific to a gauge,
giving rise to a magnetic field along the $z$ direction, and we can at
best gauge transform these solutions to obtain equivalent solutions in
a background magnetic field. The choice of the background gauge does
not permit us to obtain the free solutions in any limit as the free
solutions belong to another gauge orbit, namely the pure gauge
solutions.
\subsection{The lowest Landau level solutions}
It is previously stated that the $n=0$ solution is non-degenerate and
in this state we have only one solution available for the positive
energy and one for the negative energy. They are the $s=-1$ for the
positive energy state and $s=1$ for the negative energy state, which
is evident from Eq.\ (\ref{Usoln}) and Eq.\ (\ref{Vsoln}). Only in the
$n=0$ state the wave functions are eigenstates of $\Sigma_z$, where
$\Sigma_z = i\gamma_1 \gamma_2$, and for all other higher Landau
states the solutions do not have any definite $\Sigma_z$
eigenvalue. In actual calculations when the strength of the magnetic
field is high we require to work with the $n=0$ solutions. We can
roughly estimate the magnitude of the magnetic field suitable for the
$n=0$ approximation. Suppose we know the typical electron energy in a
system is $E$ and the magnitude of the magnetic field is ${\mathcal
B}$ from experimental observations. If it happens that $2ne{\mathcal
B}> E^2 - m^2$ for any positive value of $n$ then from the dispersion
relation in Eq.\ (\ref{En}) we see that $p_z^2$ has to be negative,
which is impossible. Consequently when ever $2e{\mathcal B}$ is
greater than the square of the typical electron energy of the system
minus the rest mass square of the electron then we have only the $n=0$
level contributing to the energy levels and only those corresponding
wave functions must be used in calculating the other details of the
system. As an example if the typical electron energy of the system is
of the order of $1{\rm MeV}$ then for magnetic field magnitude greater
than $10^{14}{\rm Gauss}$ we must only have the $n=0$ level
contributions in the energy. For lower magnitude of the magnetic field
the other Landau levels will start to contribute in the electron
energy. For a fixed energy of the electron and for very low magnetic
field magnitude we will have many possible Landau levels.
\subsection{Ortho-normality of the spinors and their completeness}
\label{orthnom}
Using the relation
\begin{eqnarray}
\int_{-\infty}^\infty  \,I_n (a) I_m (a)\,da = \sqrt{e{\mathcal B}} 
\,\,\,\delta_{n,m}\,,
\label{orthsol}
\end{eqnarray}
where $\delta_{n,m}=1$ when $n=m$ and zero otherwise and $a$ is
dimensionless we can calculate the ortho-normality of the spinors. The
ortho-normality of the spinors in the present case has to be modified
as the spinors have explicit co-ordinate dependencies. Using
Eq.~(\ref{orthsol}) it can be shown in a straight forward fashion
that,
\begin{eqnarray}
\int_{-\infty}^\infty dy\,U^\dagger_s (y,n,\vec p \omit y)
\,U_{s'} (y,m,\vec p \omit y)= 
\int_{-\infty}^\infty dy\,V^\dagger_s (y,n,\vec p \omit y)
\,V_{s'} (y,m,\vec p \omit y)
= \delta_{n,m}\delta_{s,s'} \frac{2E_n}{E_n + m}\,,
\label{uvnorm}
\end{eqnarray}
and
\begin{eqnarray}
\int_{-\infty}^\infty dy\,U^\dagger_s (y,n,\vec p \omit y)
\,V_{s'} (y,m,-\vec p \omit y)= 
\int_{-\infty}^\infty dy\,V^\dagger_s (y,n,-\vec p \omit y)
\,U_{s'} (y,m,\vec p \omit y)
= 0\,.
\label{uvonorm}
\end{eqnarray}
Except the integration over $y$ and the appearance of the Landau
levels the above relations closely resemble the corresponding
relations in free-space. The above relations fix the normalization of
the spinors. We will rederive the normalization constants of the
spinors when we quantize the theory in section \ref{aa}. 

Using now the solutions for the $U$ and the $V$ spinors from
Eqs. (\ref{Usoln}) and (\ref{Vsoln}), it is straight forward to verify
that,
\begin{eqnarray}
\int_{-\infty}^\infty dy \sum_s \Big( U_s (y,n,\vec p\omit y)^N U_s^{N\dagger} 
(y ,m,\vec p\omit y) +  V_s^N (y,n,-\vec p\omit y) 
V_s^{N\dagger} (y,m,-\vec p\omit y) \Big)
= \delta_{n,m}{\bf 1}
\label{com1}
\end{eqnarray}
Here $U^N_s (y,n,\vec p\omit y)$ and $V^N_s (y,n,-\vec p\omit y)$ are
the normalized spinors and ${\bf 1}$ is the unit $4\times 4$
matrix. If the Landau levels of the two spinors are the same then we
have,
\begin{eqnarray}
&& \sum_s \Big( U_s (y,n,\vec p\omit y) U_s^\dagger (y_\star ,n,\vec
p\omit y)
+  V_s (y,n,-\vec p\omit y) V_s^\dagger (y_\star ,n,-\vec p\omit y) \Big)
\nonumber\\*
&=& \left( 1 + {p_z^2 + 2ne{\mathcal B} \over (E_n+m)^2} \right) \times {\rm
diag} \; \Big [ I_{n-1}(\xi) I_{n-1}(\xi_\star),
I_n(\xi) I_n(\xi_\star),
I_{n-1}(\xi) I_{n-1}(\xi_\star),  I_n(\xi) I_n(\xi_\star)
\Big] \,,\nonumber\\
\label{ssumc}
\end{eqnarray}
where `diag' indicates a diagonal matrix with the specified entries,
and $\xi$ and $\xi_\star$ involve the same value of $p_x$. 
A sum over the Landau levels for spinors situated at
different $y$ co-ordinates gives,
\begin{eqnarray}
\sum_{n=0}^\infty \sum_s \Big( U^N_s (y,n,\vec p\omit y) U_s^{N\dagger} (y_\star ,n,\vec p \omit y)
+  V^N_s (y,n,-\vec p\omit y) V_s^{N\dagger} (y_\star ,n,-\vec p\omit y) 
\Big)
= \delta(y - y_\star)\,\, {\bf 1}\,,
\label{ssumc3}
\end{eqnarray}
where we have used the result of Eq.~(\ref{completeness}). The two
equations in Eq.~(\ref{com1}) and Eq.~(\ref{ssumc3}) stands for the
completeness relations for the spinors in the present case.
\section{Spin-sum of the Dirac solutions in an uniform background 
magnetic field}
\label{ssum}
In this section we derive the spin-sum $\sum_s U_s (y,n,\vec p\omit y) 
\overline U_s (y_\star,n,\vec p\omit y)$ of the solutions of the Dirac 
equation in presence of a magnetic field. The two spinors in the above
sum can have two different position coordinates in general and so
their spatial dependence is explicitly shown to be different. From the
nature of the solutions as given in Eq.\ (\ref{Usoln}) we see that
$\sum_s U_s (y,n,\vec p\omit y) \overline U_s (y_\star ,n,\vec p\omit
y)$ can be written as:
\begin{eqnarray}
P_U (y,y_\star ,n,\vec p\omit y) \equiv \sum_s U_s (y,n,\vec p\omit y) 
\overline U_s (y_\star ,n,\vec p\omit
y) = \frac{1}{E_n+m} \sum_{i,j = n-1}^n I_i(\xi) I_j(\xi_*)\,T_{i,j} 
\label{ssum1}
\end{eqnarray}
The spin-sum of the product of the spinors, $\sum_s U_s (y,n,\vec
p\omit y) \overline U_s (y_\star ,n,\vec p\omit y)$ will give rise to
a $4 \times 4$ matrix whose elements will be contain 
$I_i(\xi) I_j(\xi_*)$, where $i,\,j$ runs from $n-1,\,n$. If these
terms as $I_i(\xi) I_j(\xi_*)$ are taken as common factors then the
whole $4 \times 4$ spin-sum matrix can be represented as a sum of
terms containing the products of $I_i(\xi) I_j(\xi_*)$ times the
corresponding $4 \times 4$ matrices called $T_{i,j}$.

Using the dispersion relation $E_n^2 = p_z^2 + m^2 + 2ne{\mathcal B}$,
$T_{n,n}$ can be written as \cite{Bhattacharya:2004nj},
\begin{eqnarray}
T_{n,n}
 = 
\left( \begin{array}{ccccccc}
0 & & 0 & & 0 & & 0\\
0 & & (E_n+m) & & 0 & & p_z\\
0 & & 0 & & 0 & & 0 \\
0 & & -p_z & & 0 & & -(E_n-m)
\end{array} \right)\,.
\label{tnn}
\end{eqnarray}
In the $2\times2$ notation the above matrix can be written as,
\begin{eqnarray}
T_{n,n}
 &=& 
E_n\left( \begin{array}{ccc}
\frac12(1-\sigma_3) & & 0\\
0 & & -\frac12(1-\sigma_3)\\
\end{array} \right)
+
p_z\left( \begin{array}{ccc}
0 & & \frac12(1-\sigma_3)\\
-\frac12(1-\sigma_3)& & 0 \\
\end{array} \right)\nonumber\\
&+&
m\left( \begin{array}{ccc}
\frac12(1-\sigma_3) & & 0\\
0 & & \frac12(1-\sigma_3)\\
\end{array} \right)\,,
\label{ssum2}
\end{eqnarray}
where $\sigma_3$ is the third Pauli matrix. In the $4\times4$ notation
Eq.~(\ref{ssum2}) can be written as,
\begin{eqnarray}
T_{n,n} &=& \frac12[m(1 - \Sigma_z) + E_n(\gamma^0 + \gamma^5\gamma^3) -
p_z(\gamma^5\gamma^0 + \gamma^3)]\,,\nonumber\\
&=& \frac12[m(1 - \Sigma_z) + \rlap/p_\parallel + 
\widetilde{\rlap/p}_\parallel \gamma_5]\,,
\label{n1}
\end{eqnarray}
where $\sigma_z=i\gamma^1\gamma^2$. In the last equation
$\rlap/p_\parallel= p^0\gamma_0 +p^3\gamma_3$ and
$\widetilde{\rlap/p}_\parallel= p^0\gamma_3 +p^3\gamma_0$ and
$\gamma_5 =i\gamma^0\gamma^1\gamma^2\gamma^3$. In our case $p_0=E_n$.

In a similar way $T_{n-1,n-1}$ can be written as:
\begin{eqnarray}
T_{n-1,n-1}
 = 
\left( \begin{array}{ccccccc}
(E_n+m) & & 0 & & -p_z & & 0\\
0 & & 0 & & 0 & & 0\\
p_z & & 0 & & -(E_n-m) & & 0 \\
0 & & 0 & & 0 & & 0
\end{array} \right)\,.
\end{eqnarray}
In the $2\times2$ notation the above equation looks like,
\begin{eqnarray}
T_{n-1,n-1}
 &=& 
E_n\left( \begin{array}{ccc}
\frac12(1+\sigma_3) & & 0\\
0 & & -\frac12(1+\sigma_3)\\
\end{array} \right)
+
p_z\left( \begin{array}{ccc}
0 & & -\frac12(1+\sigma_3)\\
\frac12(1+\sigma_3)& & 0 \\
\end{array} \right)\nonumber\\
&+&
m\left( \begin{array}{ccc}
\frac12(1+\sigma_3) & & 0\\
0 & & \frac12(1+\sigma_3)\\
\end{array} \right)\,.
\end{eqnarray}
In the $4\times4$ notation the above equation becomes,
\begin{eqnarray}
T_{n-1,n-1} &=& \frac12[m(1 + \Sigma_z) + E_n(\gamma^0 - \gamma^5\gamma^3) +
p_z(\gamma^5\gamma^0 - \gamma^3)]\,,\nonumber\\
&=& \frac12[m(1 + \Sigma_z) + \rlap/p_\parallel - 
\widetilde{\rlap/p}_\parallel \gamma_5]\,.
\label{n2}
\end{eqnarray}

From the matrix multiplication in the left hand side of Eq.~(\ref{ssum1})
it can be seen that $T_{n-1,n}$ is given as,
\begin{eqnarray}
T_{n-1,n}
 = 
\sqrt{2ne{\mathcal B}}\left( \begin{array}{ccccccc}
0 & & 0 & & 0 & & 1\\
0 & & 0 & & 0 & & 0\\
0 & & -1 & & 0 & & 0 \\
0 & & 0 & & 0 & & 0
\end{array} \right)\,.
\end{eqnarray}
In the $2\times2$ notation the above equation looks like,
\begin{eqnarray}
T_{n-1,n} =
\sqrt{2ne{\mathcal B}}\left( \begin{array}{ccc}
0 & & \frac12(\sigma_1 + i\sigma_2)\\
-\frac12(\sigma_1 + i\sigma_2) & & 0
\end{array} \right)\,.
\end{eqnarray}
Here $\sigma_1$ and $\sigma_2$ are the first two Pauli matrices.
When converted back to the $4\times4$ notation the above equation
becomes,
\begin{eqnarray}
T_{n-1,n} =
-\frac12\sqrt{2ne{\mathcal B}}(\gamma_1 + i\gamma_2)\,.
\label{n3}
\end{eqnarray}

Similarly $T_{n,n-1}$ is given by,
\begin{eqnarray}
T_{n,n-1}
 = 
\sqrt{2ne{\mathcal B}}\left( \begin{array}{ccccccc}
0 & & 0 & & 0 & & 0\\
0 & & 0 & & 1 & & 0\\
0 & & 0 & & 0 & & 0 \\
-1 & & 0 & & 0 & & 0
\end{array} \right)\,.
\end{eqnarray}
In the $2\times2$ notation the above equation looks like,
\begin{eqnarray}
T_{n,n-1} =
\sqrt{2ne{\mathcal B}}\left( \begin{array}{ccc}
0 & & \frac12(\sigma_1 - i\sigma_2)\\
-\frac12(\sigma_1 - i\sigma_2) & & 0
\end{array} \right)\,,
\end{eqnarray}
which when converted back to the $4\times4$ notation becomes,
\begin{eqnarray}
T_{n-1,n} =
-\frac12\sqrt{2ne{\mathcal B}}(\gamma_1 - i\gamma_2)\,.
\label{n4}
\end{eqnarray}
Supplying the values of $T_{i,j}$s from Eq.~(\ref{n1}),
Eq.~(\ref{n2}), Eq.~(\ref{n3}) and Eq.~(\ref{n4}) to Eq.~(\ref{ssum1})
we get the result: 
\begin{eqnarray}
P_U (y,y_\star ,n,\vec p\omit y) &\equiv&
\sum_s U_s (y,n,\vec p\omit y) \overline U_s (y_\star ,n,\vec p\omit
y)= {1\over (E_n+m)} S_U (y,y_\star ,n,\vec p\omit y)\,,
\label{PU}
\end{eqnarray}
where,
\begin{eqnarray}
S_U (y,y_\star ,n,\vec p\omit y) &=& 
{1\over 2}
\bigg[\left\{ m(1+\Sigma_z) +
\rlap/p_\parallel - 
\widetilde{\rlap/p}_\parallel \gamma_5 \right\} I_{n-1}(\xi)
I_{n-1}(\xi_\star)\nonumber\\  
&+& \left\{ m(1-\Sigma_z) + \rlap/p_\parallel +
\widetilde{\rlap/p}_\parallel \gamma_5 \right\} I_n(\xi)
I_n (\xi_\star)\nonumber\\ 
&-& \sqrt{2ne{\mathcal B}} (\gamma_1 - i\gamma_2) I_n(\xi) I_{n-1}(\xi_\star) 
- \sqrt{2ne{\mathcal B}} (\gamma_1 + i\gamma_2) I_{n-1}(\xi) I_n(\xi_\star) 
\bigg] \,.\nonumber\\ 
\label{PUs}
\end{eqnarray}
Similarly, the spin sum for the $V$-spinors can
also be calculated, and we obtain:
\begin{eqnarray}
P_V (y,y_\star ,n,\vec p\omit y) &\equiv&
\sum_s V_s (y,n,\vec p\omit y) \overline V_s (y,n,\vec p\omit y) 
= {1\over (E_n+m)} S_V(y,y_\star ,n,\vec p\omit y)\,,
\label{PV}
\end{eqnarray}
where,
\begin{eqnarray}
S_V(y,y_\star ,n,\vec p\omit y) &=& 
{1\over 2}  
\Bigg[ \left\{ -m(1+\Sigma_z) +
\rlap/p_\parallel - 
\widetilde{\rlap/p}_\parallel \gamma_5 \right\} I_{n-1}(\widetilde\xi)
I_{n-1} (\widetilde\xi _\star)\nonumber\\ 
&+& \left\{ -m(1-\Sigma_z) + \rlap/p_\parallel +
\widetilde{\rlap/p}_\parallel \gamma_5 \right\} I_n(\widetilde\xi)
I_n(\widetilde\xi _\star)\nonumber\\ 
&+&\sqrt{2ne{\mathcal B}} (\gamma_1 - i\gamma_2) I_n(\widetilde\xi)
I_{n-1}(\widetilde\xi _\star) 
+ \sqrt{2ne{\mathcal B}} (\gamma_1 + i\gamma_2) I_{n-1}(\widetilde\xi)
I_n(\widetilde\xi 
_\star) \Bigg] \,.
\nonumber\\*
\label{PVs}
\end{eqnarray}
One important property of the above spin-sums is that,
\begin{eqnarray}
P_U (y,y_\star ,n,\vec p\omit y) = - P_V (y,y_\star ,n,\vec -p\omit y)\,,
\label{sspr}
\end{eqnarray}
which is similar to the result in vacuum. 
\section{Quantization of the electron field in presence of an uniform 
background magnetic field and the electron propagator}
\label{aa}
In this section we will use the spin-sum results in writing the
electron propagator in presence of an external uniform magnetic
field. But before doing so we will first write down the QED Lagrangian
for the electron in presence of a background magnetic field.

In presence of a background magnetic field we can decompose the photon
field as follows:
\begin{eqnarray}
A^\mu(x)=A^\mu_{\rm D}(x) + A^\mu_{\rm B}(x)\,,
\label{da}
\end{eqnarray}
where $A^\mu_{\rm D}(x)$ is the dynamical photon field which will be
quantized and $A^\mu_{\rm B}(x)$ is the classical background field
which gives rise to the magnetic field. If the uniform background
classical magnetic field is called ${\vec {\mathcal B}}$ then we must
have:
\begin{eqnarray}
{\vec {\mathcal B}}=\nabla \times {\vec A}_{\rm B}({\vec x})\,,
\label{mag}
\end{eqnarray}
where $A^\mu_{\rm B}(x)=(0, {\vec A}_{\rm B}({\vec x}))$. In presence
of the background magnetic field we can also write the field strength
tensor as:
\begin{eqnarray}
F^{\mu \nu}(x) = F^{\mu \nu}_{\rm D}(x) + F^{\mu \nu}_{\rm B}\,,
\label{fmunu}
\end{eqnarray}
where $F^{\mu \nu}_{\rm D}(x)=\partial^\mu A^\nu_{\rm D}(x) -
\partial^\nu A^\mu_{\rm D}(x)$ and $F^{i j}_{\rm B}= \partial^i
A^j_{\rm B}(x) - \partial^j A^i_{\rm B}(x)$ is a constant as
given in Eq.~(\ref{mag}).

The QED Lagrangian can be written as:
\begin{eqnarray}
{\mathcal L}=\overline{\psi}(i\gamma_\mu D^\mu - m)\psi -
\frac{1}{4}F^{\mu \nu}F_{\mu \nu}\,,
\label{qed1}
\end{eqnarray}
where $D^\mu=\partial^\mu - ieA^\mu$ is the covariant derivative of
the fermion fields. The QED Lagrangian can also be written as:
\begin{eqnarray}
{\mathcal L}=\overline{\psi}\left[\gamma_\mu \Pi^\mu - m\right]\psi +
e \overline{\psi}\gamma_\mu\psi A^\mu_{\rm D} - \frac{1}{4}
F^{\mu \nu}F_{\mu \nu}\,,
\label{qed2}
\end{eqnarray}
where $\Pi^\mu=i\partial^\mu + e A^\mu_{\rm B}$ is the kinetic
momentum of the fermions in presence of the background field. The
first term of the Lagrangian contains no dynamical photon dependence
but it depends upon the background magnetic field through $\Pi$ and
this part of the Lagrangian gives rise to the Hamiltonian of the
electron in presence of the magnetic field used in
Eq.~(\ref{DiracEq}). The equation of motion which we obtain from the
first term of the above Lagrangian is in fact the Dirac equation in
presence of a magnetic field which we solved in section
\ref{dsol}. Consequently the most important effect of the background
magnetic field is to modify the solutions of the Dirac equation. The
interaction term of electrons and photons remains the same as in normal
QED. The free fermionic part of the Lagrangian in Eq.~(\ref{qed2}) is
also important for the definition of the propagator of the electron
and in the next part of this section we will find out the expression
of the electron propagator in presence of the background magnetic
field. Before we calculate the electron propagator we quantize the
theory. The photons do not interact with the magnetic field and
consequently their quantization procedure is the same as in normal
QED.
\subsection{Quantization of the electron field}
Since we have found the solutions to the Dirac equation, we can now
use them to construct the fermion field operator in the second
quantized version. For this, we write
\begin{eqnarray}
\psi(X) = \sum_{s=\pm} \sum_{n=0}^\infty \int {dp_x \, dp_z \over 2\pi D}
\left[ f_s (n,\vec p\omit y) e^{-ip\cdot X {\omit y}} U_s (y,n,\vec
p\omit y) +
\widehat f_s^\dagger (n,\vec p\omit y) e^{ip\cdot X {\omit y}} V_s
(y,n,\vec p\omit y) \right] \,.
\label{2ndquant}
\end{eqnarray}
Here, $f_s(n,\vec p\omit y)$ is the annihilation operator for the
fermion, and $\widehat f_s^\dagger(n,\vec p\omit y)$ is the creation
operator for the antifermion in the $n$-th Landau level with given
values of $p_x$ and $p_z$. It is to be noted that the wave-functions
of the electron used in Eq.~(\ref{2ndquant}) are not free-particle
solutions and they never tend to the free-particle solutions in any
limit. As the the positive and negative frequency parts of the
solutions are as free-particles so the notion of a particle and
anti-particle is unambiguous in the present circumstance.  The creation
and annihilation operators satisfy the anti-commutation relations
\begin{eqnarray}
\left[ f_s (n,\vec p\omit y), f_{s'}^\dagger (n',\vec p'\omit y)
\right]_+ =
\delta_{ss'} \delta_{nn'} \delta(p_x-p'_x) \delta (p_z - p'_z) \,,
\label{freln}
\end{eqnarray}
and a similar one with the operators $\widehat f$ and $\widehat
f^\dagger$, all other anti-commutators being zero. The quantity $D$
appearing in Eq.\ (\ref{2ndquant}) depends on the normalization of the
spinor solutions, and in this section we will rederive the
normalization of the spinors calculated in subsection
\ref{orthnom}. The factor of $2\pi$ multiplying $D$ helps to associate
$D$ with the normalization constant found in subsection \ref{orthnom}.
Once we have chosen the spinor normalization, the factor $D$ appearing
in Eq.\ (\ref{2ndquant}) is however fixed, and it can be determined
from the equal time anti-commutation relation
\begin{eqnarray}
\left[ \psi(X), \psi^\dagger(X_\star) \right]_+ = \delta^3 (\vec X - \vec
X_\star) \,.
\label{anticomm}
\end{eqnarray}
Plugging in the expression given in Eq.\ (\ref{2ndquant}) to the left
side of this equation and using the anti-commutation relations of Eq.\
(\ref{freln}), we obtain
\begin{eqnarray}
\left[ \psi(X), \psi^\dagger(X_\star) \right]_+ = \sum_{s} \sum_{n} \int
{dp_x \, dp_z \over (2\pi D)^2} &&
\Big( e^{-ip_x(x-x_\star)} e^{-ip_z(z-z_\star)}
U_s (y,n,\vec p\omit y) U_s^\dagger (y_\star ,n,\vec p\omit y)
\nonumber\\*
&& +  e^{ip_x(x-x_\star)} e^{ip_z(z-z_\star)}
V_s (y,n,\vec p\omit y) V_s^\dagger (y_\star ,n,\vec p\omit y) \Big)
\,.\nonumber\\
\end{eqnarray}
Changing the signs of the dummy integration variables $p_x$ and $p_z$
in the second term, we can rewrite it as
\begin{eqnarray}
\left[ \psi(X), \psi^\dagger(X_\star) \right]_+ = \sum_{s} \sum_{n} \int
{dp_x \, dp_z \over (2\pi D)^2} && e^{-ip_x(x-x_\star)}
e^{-ip_z(z-z_\star)} \Big(
U_s (y,n,\vec p\omit y) U_s^\dagger (y_\star ,n,\vec p\omit y)
\nonumber\\*
&& +  V_s (y,n,-\vec p\omit y) V_s^\dagger (y_\star ,n,-\vec p\omit y)
\Big) \,.
\label{anticomm1}
\end{eqnarray}
At this stage, we can perform the sum over $n$ in Eq.\
(\ref{anticomm1}) using Eq.~(\ref{ssumc}) and Eq.\
(\ref{completeness}) which gives the $\delta$-function of the
$y$-coordinate and perform the integrations over $p_x$ and $p_z$ to
recover the $\delta$-functions for the other two coordinates as well,
provided
\begin{eqnarray}
{2E_n \over E_n+m} \; {1\over (2\pi D)^2} = {1\over (2\pi)^2} \,.
\end{eqnarray}
In this way we get back the same value of the
normalization of the spinors which we obtained in subsection
\ref{orthnom}. Putting the solution for $D$, we can rewrite Eq.\
(\ref{2ndquant}) as
\begin{eqnarray}
\psi(X) &=& \sum_{s=\pm} \sum_{n=0}^\infty \int {dp_x \, dp_z \over
2\pi} \sqrt {E_n+m \over 2E_n} \nonumber\\* && \times
\left[ f_s (n,\vec p\omit y) e^{-ip\cdot X {\omit y}} U_s (y,n,\vec
p\omit y) +
\widehat f_s^\dagger (n,\vec p\omit y) e^{ip\cdot X {\omit y}} V_s
(y,n,\vec p\omit y) \right] \,.
\label{psi}
\end{eqnarray}

The one-fermion states are defined as
\begin{eqnarray}
\left| n,\vec p\omit y, s \right> = C f_s^\dagger (n,\vec p\omit y) \left|
0 \right> \,.
\end{eqnarray}
The normalization constant $C$ is determined by the condition that the
one-particle states should be orthonormal. For this, we need to define
the theory in a finite but large region whose dimensions are  $L_x$,
$L_y$ and $L_z$ along the three spatial axes. This gives
\begin{eqnarray}
C = {2\pi \over \sqrt{L_x L_z}} \,.
\end{eqnarray}
Next we calculate the electron propagator in presence of an uniform
background magnetic field.
\subsection{The electron propagator}
The electron propagator is given by,
\begin{eqnarray}
i S^B_F (X - X_\star)_{\alpha \beta} &=& \langle 0|T\{\psi(X)_\alpha 
\overline{\psi}(X_\star)_\beta\}|0\rangle \nonumber\\
&=& \theta(t - t_\star) \langle 0|\psi(X)_\alpha \overline{\psi}(X_\star)_
\beta |0\rangle - \theta(t_\star - t) \langle 0|\overline{\psi}
(X_\star)_\beta  \psi(X)_\alpha|0\rangle\,,\nonumber\\
\label{eprop}
\end{eqnarray}
where $T$ is the time-ordered product and $\theta(\lambda)$ is the
step-function which is unity when $\lambda \ge 0$ and zero other
wise. The step function can be represented in an integral form as:
\begin{eqnarray}
\theta(\lambda) = i \int_{-\infty}^\infty \frac{d\omega}{2\pi}
\frac{e^{-i\lambda\omega}}{\omega - i\epsilon}\,,
\label{sf} 
\end{eqnarray}
where $\epsilon$ is an infinitesimal parameter. Using Eq.~(\ref{psi})
we can write,
\begin{eqnarray}
\langle 0|\psi(X)_\alpha \overline{\psi}(X_\star)_\beta \}|0\rangle
&=& \sum_{s=\pm} \sum_{n=0}^\infty \int {dp_x \, dp_z \over
(2\pi)^2} \left( E_n+m \over 2E_n \right)
e^{-ip\cdot (X {\omit y} - X_\star {\omit y})}\nonumber\\
&\times& U_{s,\alpha} (y,n,\vec p\omit y)
{\overline U}_{s, \beta} (y_\star,n,\vec p\omit y)\,,\nonumber\\
\end{eqnarray}
and using Eq.~(\ref{PU}) and suppressing the spinor indices the above
equation can also be written as,
\begin{eqnarray}
\langle 0|\psi(X) \overline{\psi}(X_\star)|0\rangle
= \sum_{n=0}^\infty \int {dp_x \, dp_z \over
(2\pi)^2} \left( E_n+m \over 2E_n \right)
e^{-ip\cdot (X {\omit y} - X_\star {\omit y})} 
P_U(y, y_\star,n,\vec p\omit y)\,.
\end{eqnarray}
In a similar way it can be shown that,
\begin{eqnarray}
\langle 0|\overline{\psi}(X_\star) \psi(X)|0\rangle
= \sum_{n=0}^\infty \int {dp_x \, dp_z \over
(2\pi)^2} \left( E_n+m \over 2E_n \right)
e^{ip\cdot (X {\omit y} - X_\star {\omit y})}
P_V(y, y_\star,n,\vec p\omit y)\,,
\end{eqnarray}
where $P_V$ is given in Eq.~(\ref{PV}). Using the above results in
Eq.~(\ref{eprop}) and utilizing the form of the $\theta$-function
in Eq.~(\ref{sf}) we can write,
\begin{eqnarray}
& &i S^B_F (X - X_\star)= i\sum_{n=0}^\infty \int {dp_x \, dp_z \, 
d\omega \over (2\pi)^3} \left( E_n+m \over 2E_n \right) \nonumber\\
&\times& \left[\frac{e^{-i\omega(t - t_\star) - ip\cdot (X {\omit y} - X_\star 
{\omit y})}}{\omega - i\epsilon} P_U(y, y_\star,n,\vec p\omit y) -
\frac{e^{i\omega(t - t_\star) + ip\cdot (X {\omit y} - X_\star
{\omit y})}}{\omega - i\epsilon} P_V(y, y_\star,n,\vec p\omit y)\right]\,.
\end{eqnarray}
Changing the signs of $p_x$ and $p_z$ in the second term of the
integrand and using Eq.~(\ref{sspr}) we get,
\begin{eqnarray}
i S^B_F (X - X_\star) &=& i\sum_{n=0}^\infty \int {dp_x \, dp_z \, 
d\omega \over
(2\pi)^3} e^{i{\bf p}\cdot ({\bf X} {\omit y}- {\bf X}_\star{\omit y})}
 \left( E_n+m \over 2E_n \right)P_U(y, y_\star,n,\vec p\omit y) \nonumber\\
&\times& \left[\frac{e^{-i(\omega + E_n)(t - t_\star)}}{\omega - i\epsilon} +
\frac{e^{i(\omega + E_n)(t - t_\star)}}{\omega - i\epsilon}\right]\,.
\end{eqnarray}
Now appropriately doing the integration over $\omega$ we get,
\begin{eqnarray}
i S^B_F (X - X_\star) &=& i\sum_{n=0}^\infty \int {dp_x \, dp_z \, dp_0 \over
(2\pi)^3} e^{-i{p}\cdot ({ X} {\omit y}- {X}_\star {\omit y})}
\frac{S_U(y, y_\star,n,\vec p\omit y)}{p_0^2 - p_z^2 - m^2 - 2ne{\cal B} - 
i\epsilon}\,,\nonumber\\
&=&i\sum_{n=0}^\infty \int {dp_x \, dp_z \, dp_0 \over
(2\pi)^3} e^{-i{p}\cdot ({ X} {\omit y}- {X}_\star {\omit y})}
\frac{S_U(y, y_\star,n,\vec p\omit y)}{p_\parallel^2 - m^2 - 2ne{\cal B} -
i\epsilon}\,, 
\label{propex}
\end{eqnarray}
where $S_U(y, y_\star,n,\vec p\omit y)$ is given by Eq.~(\ref{PUs})
and $p_\parallel^2 = p_0^2 - p_z^2$. It is to be noted that the pole
of the propagator is now dependent on the Landau levels as it should
be in an uniform background magnetic field. The form of the propagator
suggests that it is not translation invariant and so it cannot be
written down completely in Fourier space. 
\section{A discussion on gauge dependence}
As we have solved the Dirac equation in presence of a uniform
background magnetic field using a particular gauge, as given in
Eq.~(\ref{GA}) with $b=0$, the solutions are dependent on the gauge
choice. The spinor solutions are themselves not physical observables
and so they can be gauge dependent. But not all the results discussed
in this article are gauge dependent. The energy of the electron as
given in Eq.\ (\ref{E}) is not a gauge dependent quantity, any gauge
we choose we will get the same dispersion relation of the
electrons. The special forms of the ortho-normality relations as given
in section \ref{orthnom} are gauge dependent as the results contain
the functions which has $p_x$ which is not a gauge invariant
quantity. The spin-sum also depends on the particular gauge we work in
and the above results will be different if we had chosen another gauge
to represent the magnetic field. But actual calculations yielding
physical quantities like scattering cross-section or decay rates must
be independent of the choice of the background gauge field. We can see
the gauge invariance of the physical quantities in a heuristic way. If
we had chosen the gauge specified in Eq.~(\ref{GB}) with $c=0$ instead
of the gauge in Eq.~(\ref{GA}) with $b=0$ then the solutions of the
Dirac equations as specified in Eq.~(\ref{Usoln}) and
Eq.~(\ref{Vsoln}) should have been the same except all the $y$ should
be replaced by $x$ and $p_x$ should be replaced by $p_y$ inside the
spinors and the free-particle part should contain $e^{ip\cdot X {\omit
x}}$ instead of $e^{ip\cdot X {\omit y}}$. A similar replacement
should yield the new spin-sums and the propagator. Consequently the
quantities calculated in these two gauges differ by the way we name
the $x$ and $y$ coordinate axes. But in calculations of scattering
cross-sections and decay-rates we always have integrations over $x, y,
z$ coordinates at each vertex and consequently the end results will
not depend upon which gauge we started with.
 
The above discussion highlights the fact that most of the quantities
calculated in this article using the exact solutions in presence of
the magnetic field rely heavily on our choice of the vector
potential. All the solutions of the Dirac equation in presence of a
uniform magnetic field along the $z$ direction obtained by using
various vector potentials will be different but are related by gauge
transformations. It is to be noted that the free Dirac solutions can
also be gauge rotated where the gauge fields are pure gauge
configurations. As there is no connection between the gauge
configurations giving rise to a magnetic field along the $z$ direction
and pure gauge fields so we do not get back the free Dirac solutions
as a limit of the exact solutions in a magnetic field.
\section{Conclusion}
In this article we solved the Dirac equation in presence of a
background uniform magnetic field specified by a particular vector
potential. The dispersion relation of the electron is seen to change
from its form in the vacuum and we see the emergence of Landau levels
designating the quantized nature of the transverse motion of the
electrons. The solutions of the Dirac equation are dependent on the
Landau levels, the energy of the electron is seen to be degenerate
except the lowest Landau level energy. It is seen that there is no way
to get back the free Dirac solutions from the exact solutions in
presence of the magnetic field by letting the field strength to go to
zero in the solutions, a fact which is related to the gauge invariance
of the system. Using the appropriate spinors in a magnetic field the
ortho-normality and completeness of the spinors were worked out in
section \ref{orthnom}, which closely resembles the corresponding
results in vacuum.  The spin-sum of the solutions are derived
explicitly using the exact solutions of the Dirac equation in a
magnetic field. The theory is quantized and with the quantum field
operators the propagator of the electron in presence of an uniform
background magnetic field is calculated. Some thing similar to our
derivation of the electron propagator was presented in
\cite{Kobayashi:1983dt} where the authors worked in the chiral
representation of the Dirac gamma matrices. But the presentation of
the expression of the propagator was not compact and nor the authors
in \cite{Kobayashi:1983dt} calculate the spin-sum explicitly. As most
of the quantities calculated in this article depend on the choice of
the vector potential giving rise to the magnetic field so the gauge
invariance of the calculations become less transparent. In the
penultimate section we discuss about the gauge invariance of the
calculations in presence of a magnetic field and show that although
the spin-sums and propagators may not be gauge invariant but physical
quantities like scattering-cross sections and decay rates of
elementary particles in presence of a magnetic field can be gauge
invariant.

\end{document}